\documentclass[a4paper,twoside]{article}
\pdfoutput=1
\usepackage{epsfig}
\usepackage{subcaption}
\usepackage{calc}
\usepackage{amssymb}
\usepackage{amstext}
\usepackage{amsmath}
\usepackage{amsthm}
\usepackage{multicol}
\usepackage{pslatex}
\usepackage{apalike}
\usepackage{url}
\usepackage{SCITEPRESS} 
\usepackage{dblfloatfix}
\usepackage{graphicx}
\usepackage{hyperref}
\hypersetup{colorlinks,allcolors=black}
\usepackage{slashbox}
\usepackage{footmisc}
\usepackage{cite}
\usepackage{amsmath,amssymb,amsfonts}
\usepackage{multirow,hhline,graphicx,array}
\usepackage{graphicx}
\usepackage{tabularx}
\usepackage{textcomp}
\usepackage{xcolor}
\usepackage{import}
\usepackage{soul}
\usepackage{rotating}
\usepackage{adjustbox}
\usepackage{multirow}
\usepackage{array}
\usepackage{comment}
\usepackage{makecell}
\usepackage{float}
\setcellgapes{4pt}
\usepackage{multicol,lipsum}
\usepackage{lipsum}% http://ctan.org/pkg/lipsum
\usepackage{algorithm}% http://ctan.org/pkg/algorithm
\usepackage{algpseudocode}% http://ctan.org/pkg/algorithmicx
\usepackage{caption}% http://ctan.org/pkg/caption

\def\BibTeX{{\rm B\kern-.05em{\sc i\kern-.025em b}\kern-.08em
    T\kern-.1667em\lower.7ex\hbox{E}\kern-.125emX}}

\makeatletter
\let\OldStatex\Statex
\renewcommand{\Statex}[1][3]{%
  \setlength\@tempdima{\algorithmicindent}%
  \OldStatex\hskip\dimexpr#1\@tempdima\relax}
\makeatother
\newcommand*\rot{\rotatebox[origin=c]{90}}

\newcommand{\todo}[1]{}
\renewcommand{\todo}[1]{{\color{red} TODO: {#1}}}
    
\makeatletter
\newcommand{\thickhline}{%
	\noalign {\ifnum 0=`}\fi \hrule height 1pt
	\futurelet \reserved@a \@xhline
}
\newcolumntype{"}{@{\hskip\tabcolsep\vrule width 1pt\hskip\tabcolsep}}

\newcommand{\newlineauthors}{%
  \end{@IEEEauthorhalign}\hfill\mbox{}\par
  \mbox{}\hfill\begin{@IEEEauthorhalign}
}

\makeatother

\begin{document}

\title{RCM: Requirement Capturing Model for Automated Requirements Formalisation} 

%\begin{comment}

\author{\authorname{Aya Zaki-Ismail\sup{1}%\orcidAuthor{0000-0000-0000-0000}
, Mohamed Osama\sup{1}%\orcidAuthor{0000-0000-0000-0000}, 
,Mohamed Abdelrazek\sup{1}%\orcidAuthor{0000-0000-0000-0000} 
, John Grundy\sup{2}%\orcidAuthor{0000-0000-0000-0000} 
and Amani Ibrahim\sup{1}%\orcidAuthor{0000-0000-0000-0000}
}
\affiliation{\sup{1}Information Technology Institute, Deakin University, 3125 Burwood Hwy, VIC, Australia}
\affiliation{\sup{2}Information Technology Institute, Monash University, 3800 Wellington Rd, VIC, Australia}
\email{\{amohamedzakiism, mdarweish, mohamed.abdelrazek, amani.ibrahim\}@deakin.edu.au, john.grundy@monash.edu}
}
%\end{comment}


\abstract{
%Mission critical systems, e.g. vehicles and medical equipment, need to be formally verified using well-established formal verification techniques. These systems are becoming dramatically larger in size and complexity. 
Most existing automated requirements formalisation techniques require system engineers to (re)write their requirements using a set of predefined requirement templates with a fixed structure and known semantics to simplify the formalisation process. However, these techniques require understanding and memorising requirement templates, which are usually fixed format, limit requirements captured, and do not allow capture of more diverse requirements. To address these limitations, we need a reference model that captures key requirement details regardless of their structure, format or order. Then, using NLP techniques we can transform textual requirements into the reference model. Finally, using a suite of transformation rules we can then convert these requirements into formal notations. In this paper, we introduce the first and key step in this process, a Requirement Capturing Model (RCM) - as a reference model - to model the key elements of a system requirement regardless of their format, or order. We evaluated the robustness of the RCM model compared to 15 existing requirements representation approaches and a benchmark of 162 requirements. Our evaluation shows that RCM breakdowns support a wider range of requirements formats compared to the existing approaches. We also implemented a suite of transformation rules that transforms RCM-based requirements into temporal logic(s). In the future, we will develop NLP-based RCM extraction technique to provide end-to-end solution. %In the future, we will introduce an NLP-based approach to transform textual requirements into RCM-based requirements model.
}
\onecolumn \maketitle %\normalsize %\setcounter{footnote}{0} \vfill

%\import{./}{abstract.tex}

\section{\uppercase{Introduction}}

%The world is observing a wide-adoption of Internet-of-Things and Cyber-Physical systems to assist and automate many mission critical tasks in many areas - e.g. automotive industry, robotics, smart infrastructure, smart homes and smart cities, etc. Errors or crashes in such systems can lead to disasters \cite{sternudd2011unambiguous}. Prevention of such outcomes mandates stringent monitoring of the quality of the system during the software and system development life-cycle  \cite{sternudd2011unambiguous} \cite{CFGTh} \cite{konrad2005realTime}. Many of the existing standards \cite{sternudd2011unambiguous, iso201126262, IEEE830} mandate formal verification of these systems or the key components.
 
Formal verification techniques requires system requirements to be expressed in formal notations \cite{buzhinsky2019formalization}. However, the majority of critical system requirements are still predominantly written in informal notations (textual or natural languages - NL), which are inherently ambiguous  and have incomplete syntax and semantics \cite{lucio2017EARSAnalysis, CFGTh}. To automate the formalisation process, several bodies of work within the literature focused on proposing pre-defined requirement templates, patterns \cite{BTCTh}, boilerplates \cite{mavin2009EARS}, and structured control English \cite{ACE}, to express one system requirement sentence while eliminating the ambiguities. Such templates have complete syntax to ensure the feasibility of transforming textual requirements into formal notations using a suite of manually crafted, template-specific transformation rules (e.g., \cite{yan2015formal}). However, some of the predefined templates are domain dependent and are hard to generalise \cite{rupp2009rupboilerplates}, or can only capture limited subsets of requirements structures \cite{ACE}. In addition, most existing formalisation algorithms are customized for transforming system requirements to one target formal language. Thus, a need to transform the same requirements into different formal languages mandates significant rework of the formalisation algorithm.% \todo{JG - agree - cite papers saying this}.

Complementary to this research direction, instead of considering introducing new sentence-based templates covering a wider range of requirements and complicating the requirements specification process, we introduce a Requirement Capturing Model (RCM), as a reference model that defines the key properties that make up a system behavioral requirement sentence, \emph{regardless of the syntactic structure of these properties, lexical-words, or their order}. RCM separates the writing styles (format and structure) from the abstract requirement properties and the formal notations. Our new RCM model thus enables us to: (1) represent a much wider range of requirements that have differing count, order or types of properties, by identifying the specific properties in the input requirement sentence to generic RCM defined properties; (2) specify requirements in a wide variety of different formats, extremely useful to avoid re-writing existing requirements; (3) formalize requirements into different formal notations through mapping RCM properties to those of the target formal notation; and (4) enable use of NLP-based requirements extraction techniques to transform textual requirements into the RCM-based requirements model. with the key elements to be extracted now clearly defined and known. 
Our key contributions in this paper are:

%Complementary to this research direction, we target a new requirements formalisation paradigm aiming to increase the flexibility in representing diverse requirements formats and covering a wider range. This is achieved by introducing a well-defined intermediate representation -- \textbf{a reference model} -- that defines the key requirement properties that exist in an input requirement, \emph{regardless of the natural language format, order, or structure of these properties}. Using this reference model we can develop Natural Language Processing (NLP) techniques to automatically extract these requirement properties from the input requirement. Finally, we can develop a suite of transformation rules to transform system requirements represented in this reference model into necessary formal notations. Furthermore, the RCM can be used to generate precise requirement documents and different sorts of requirement models using suites of transformation rules.  
%In this paper, we focus on the first and third step in this formalisation paradigm. 

\begin{itemize}
    \item Introduce RCM as a reference model and intermediate representation between informal and formal notations. RCM was developed based on extensive review of existing requirements templates, patterns, CNL, etc in the literature, with a view to support automatic transformation into formal notations.
    \item A suite of transformation rules from RCM to Metric Temporal Logic (MTL), to demonstrate how an RCM-based requirements model can be transformed into formal notations.   
    %\item A large, representative dataset of 162 \todo{behavioral requirements for critical systems} synthesized from the literature for the evaluation of natural language formalisation techniques.
    \item Evaluation of the representation power of RCM by comparing it to 15 other existing approaches using 162 behavioral requirements for critical systems synthesized from the literature. We provide the RCM representation and corresponding automatically generated (MTL and CTL) formal notations for each of these requirements. %We are working on new requirements extraction technique to extract RCM from textual requirements.%, and also a suite of transformation rules from RCM to Formal Notations.
\end{itemize}

%There are two other necessary building blocks in order to offer a complete end-to-end requirements formalisation toolset including: 1) RCM-bassed requirements quality checking tool, and 2) an NLP-based tool to automatically extract RCM model constructs from input textual requirements. These two tools are currently under evaluation and are outside the scope of this paper. 

%This would allow for an end-to-end requirements formalisation toolset. 

%The rest of this paper is organized as follows: Section \ref{motivation} provides motivating scenario. Section \ref{relatedWork}, illustrates existing work. Section \ref{RCMOverview}, proposes the RCM model highlighting: the meta-model, formal mapping and formal semantic augmentation. Section \ref{RCMFormalization} shows requirements transformation into formal notations. In section \ref{evaluation}, we evaluate the RCM approach. Section \ref{conclusion}, concludes the paper and discuss future work.

\begin{comment}
 Section \ref{motivation} provides a motivating example and outlines the main challenges to process textual requirements.
 
 Many of these critical systems encompass time-dependent components that have special characteristics (e.g., (1)integrated system, (2) performing tasks within specified time frame, if the brake system of a car exceeds the time limit, it may cause accidents. (3) reacting to external stimuli and reacting accordingly like GPS, etc). 
    
\end{comment}
   
\section{\uppercase{Motivation}} \label{motivation}
Jen is a system engineer working for an automotive company. She wants to specify the requirements of one of the system modules - a small excerpt is shown in Table \ref{tab1:Req} - while making sure that these requirements can be easily transformed into formal notations as a mandatory compliance requirement. Jen decided to check the existing requirement specification techniques in the literature to choose which one covers most of her requirements. Jen researched existing requirements formalisation techniques, see the related work section for these techniques, and outlined her trials to use these techniques to model her requirements after rephrasing some of her requirements to suit existing templates. 

\begin{table}[htbp]
	\caption{Examples of critical System Requirements and approaches to represent}
	
	\scalebox{0.85}{
		\begin{tabular}{|p{8.5cm}|}
		%	\hline
		%	RQ1: If the fuel amount is above \textbf{threshold (config:Fuel\_min\_threshold)}, the fuel level is set to HIGH. If \textbf{the above threshold} exceeds the fuel amount, the fuel level is set to LOW and the fuel light shall plink.\\
	%		\hline
	%		RQ2: If the engine oil amount is above \textbf{threshold (config:Eng\_oil\_min\_threshold)}, the engine oil level is set to High. If the engine oil amount is below \textbf{the above threshold}, the level is set to low and the engine oil shall plink. \\
			\hline
			\textbf{RQ1:} {R\_STATUS} shall indicate the rain sensor. {It} shall be ON, when the external environment is raining. \\\\
			
			\textbf{Techniques:} Universal pattern \cite{teige2016universal},
			Structured English \cite{konrad2005realTime}, 
			Rup's boilerplates \cite{rupp2009rupboilerplates}, ACE\cite{ACE}, EARS\cite{mavin2009EARS},  CFG\cite{CFGTh} and BTC\cite{BTCTh}\\
			
			\hline
			\textbf{RQ2:} When \underline{the external environment rains} for 1 minute, the wipers shall be activated within 30 seconds until the rain sensor equals OFF.\\\\

			\textbf{Techniques:} Universal pattern \cite{teige2016universal} and BTC\cite{BTCTh} \\
			
			\hline
			
			\textbf{RQ3:} While the wipers are active, the wipers speed shall be readjusted every 20 seconds.\\\\
			%If a crash is detected, the emergency light shall be activated for 2 seconds every 30 seconds.\\
			\textbf{Techniques:} Structured English \cite{konrad2005realTime} \\
			\hline
		\end{tabular}
		\label{tab1:Req}
	}
\end{table}

Jen found that none of the existing techniques she found can be used to cover all her requirements. She then had to learn and use all these templates and have these tools all running. Furthermore, Jen found that the majority of these solutions rely on pre-defined formats and structure of requirements boilerplates. This mandates (1) a fixed order of requirement components/sub-components, (2) a fixed English-syntax for a specific component/sub-component, (3) a fixed/small set of English verbs or other lexical words. Thus, Jen needs to rewrite her requirements to confirm the defined format which puts more overhead on her especially if the defined formats are limited and cannot be extended to new scenarios.  

Taking into consideration all combinations of styling, ordering, and omission/existence of different requirements model properties will increase the size of the defined formats. Consequently this will increase the complexity of using them by system engineers and the complexity of the parsing algorithms needed to transform them to formal models. Furthermore, most existing formalization techniques apply on-the fly transformation on the given structured requirement sentences to generate formal notations. These transformations are hard-coded or tightly customized according to the target formal notation properties and formats. It would be much more useful if the common parts are computed once and transformed to intended notations as needed. %More importantly, system engineers do not get to see the process and it is still tricky to judge if the generated notations reflect the input requirements. 

%Having a user-friendly middle (intermediate) step where they can see the extracted components before formalising these. This requires a rich intermediate model that supports isolating the requirement structure and formats, key components and sub-components in the requirements, and the corresponding target formal notations would streamline the formalisation process.

\section{Related Work} \label{relatedWork}
%We outline key efforts to formalise natural language requirements. This also helps identifying the key properties (constructs or components/sub-components) that could exist in an input requirement. This helps ensuring our intermediate representation is comprehensive. 
%This section serves two objectives: outline the key efforts in the area and their limitation when it comes to formalising requirements, but more importantly deriving the key components of the requirements capturing model that need to be addressed in the intermediate reference model.

Many requirements formalisation approaches assume requirements are specified in a constrained natural language (CNL) with specific style, format and structure to be able to transform into formal notations - e.g. \cite{ghosh2016arsenal, nelken1996automatic, michael2001natural, holt1999semantically, ambriola1997processing, sturla2017two, pease2010controlled}. These CNL are meant to avoid natural language related quality problems (e.g., ambiguity inconsistency, etc.) and increase the viability of automating the formalisation process.

CNL is a restricted form of NL especially created for writing technical documents as defined in \cite{kittredge2003sublanguages} with the aim to reduce/avoid NL problems (e.g., ambiguity inconsistency, .etc). CNL typically has a defined sub-set of NL grammar, lexicon and/or sentence structure \cite{kuhn2014survey}. Different forms of CNL are also provided as a reliable solution for requirements representation. Fuchs et al. \cite{ACE} propose Attempto Controlled English (ACE) with a restricted list of verbs, nouns and adjectives for the requirement set in addition to restrictions on the structure of the sentence. ACE can be transformed into Prolog. ACE can handle requirements with condition and action components. Multiple CNLs are proposed later inspired by ACE  (e.g., Atomate language \cite{van2010atomate}, PENG\cite{schwitter2002english}) for formal generation purposes and for other purposes (e.g., BioQuery-CNL \cite{erdem2009transforming}, OWL ACE \cite{kaljurand2006bidirectional}).

Similarly to ACE, Scott and Cook\cite{scott2004context} presented Context Free Grammars (CFGs) for requirement specification. Although the format of the requirement components is more limited than ACE with additional restrictions on words, it covers time-related properties. Yan et al. \cite{yan2015formal} presented a more flexible CNL with constraints on the word set such that, a clause should contain (1) single word noun as a subject and a verb predicate with one of the following formats "verb $|$ be+(gerund$|$participle) $|$ be+complement", (2) the complement should be adjective or adverbial word, (3) prepositional phrases are not allowed except "in + time point" at the end of the clause. The CNL does not consider time information except pre-elapsed time.

Boilerplates are also widely used. These provide a fixed syntax and lexical words with replaceable attributes. Boilerplates are more limited than CNL and require adaptation to different domains. In \cite{rupp2009rupboilerplates}, a constrained RUP’s boilerplate is provided which can handle a limited range of requirements.  EARS \cite{mavin2009EARS} boilerplates are less restricted and can support a wider range of requirements. Esser et al. \cite{esser2007obtaining} proposed a suite of requirement templates (TBNLS) with support mapping to propositional logic with temporal relations. For validating the conformity of the written requirement and the boilerplate, authors in \cite{arora2013rubricChcking1, arora2014requirementChcking2} provide checking techniques. 

Requirement patterns provide a more flexible solution. However, When a new requirement structure is added, a new pattern should be created for it, which increases the size of the patterns set. In \cite{teige2016universal} a universal pattern was presented to support many requirements formats (trigger, then action). They then introduced additional time-based kernel patterns in \cite{BTCTh}. Although these patterns cover many requirement properties, they do not still cover the possible combinations of the supported properties eligible to one requirement specification. In addition, the approach lack complex time properties - e.g. In-between-time and pre-elapsed-time properties. Dwyer et al. \cite{dwyer1999patterns} proposed several patterns applicable for non-real-time requirement specifications. These patterns are categorized into two major groups: occurrence patterns and order patterns, while considering scopes (e.g., globally, before R, after R) for a given specification pattern. The work is extended later in \cite{konrad2005realTime} to cope with real-time requirement specifications. The real time patterns considers versions of the pre-elapsed-time, in-between-time and valid-time information for the action component.

Event-Condition-Action (ECA) was initially proposed in active databases area to express behavioral requirements. ECA became widely used by several researchers in diffident areas. An ECA rule assumes that when an event E occurs, the condition C will be evaluated, and if true, the action A will be executed. ECA notations have been extended to capture time information \cite{qiao2007developing}. However, ECA rules do not support (\textit{e.g.}, factual rules), and do not consider scopes for action and the time notations apply on events. 

Despite extensive research and industrial use, all of these alternatives including CNL, patterns, boilerplates and ECA have downsides including: (1) users need guidance on how to phrase requirements in terms of CNL; (2) expressiveness is reduced by the available subset of natural language used; and (3) they are restricted to certain domains or pre-defined subsets of requirements properties.  RCM subsumes and builds on the key properties (components) introduced in all of these techniques, but with enough detail and flexibility to facilitate the requirements formalisation as discussed below.

\begin{table*}[!b]
\caption{A list of identified requirement properties from existing approaches}
\begin{center}
\scalebox{0.9}{
\small
%\begin{tabular}{|p{.43cm}|p{.5cm}|p{.2cm}|p{.2cm}|p{.2cm}|p{.5cm}|p{.2cm}|p{.2cm}|p{.5cm} |p{.2cm}|p{.2cm}|p{.5cm}|p{.2cm}|p{.5cm} |p{.2cm}|p{.5cm}|p{.2cm}|p{.5cm}|p{.2cm}|p{.7cm}| }

\begin{tabular}{|m{1em}|l|p{14cm}|}

\hline
\multicolumn{2}{|l|}{\textbf{Property}} & \textbf{Description}\\\hline
\multirow{4}{2cm}{}

    &\textbf{Trigger} & \begin{tabular}{p{14cm}} is an event that initiates action(s) (e.g., "when the system halts" in Fig.\ref{fig:reqEX}). This component type is ubiquitous throughout the requirements of most critical systems. \end{tabular}\\ \cline{2-3}

    & \textbf{Condition} &  \begin{tabular}{p{14cm}}is a constraint that should be satisfied to allow a specific system action(s) to happen (\textit{e.g.,} "if X is ON" in Fig.\ref{fig:reqEX}). In contrast to triggers, the satisfaction of the condition should be checked explicitly by the system. The system is not concerned with "when the constraint is satisfied" but with "is the constraint satisfied or not at the checking time" to execute the action (e.g., in the previous example "X" might remain "ON" for a while and have no effect on the system until checked for. \end{tabular} \\ \cline{2-3}
    
   \multirow{2}{2cm}{\rot{\textbf{Component}}} & \textbf{Action} & \begin{tabular}{p{14cm}} is a task that should be accomplished by the system in response to triggers and/or constrained by conditions (e.g., "M should be set to TRUE" in Fig.\ref{fig:reqEX}). In case that, a primitive requirement consists of an action component only, it would be marked as \textbf{a factual rule} expressing factual information about the system (e.g., \textit{The duration of a flashing cycle is 1 second} \cite{houdek2013system}). \end{tabular}
    \\ \cline{2-3}
    &  \textbf{Req-scope} & \begin{tabular}{p{14cm}} determines the context under which (i) "condition(s) and trigger(s)" can be valid -- called a pre-conditional scope as it is linked to the condition or trigger; and (ii) "action(s)" can occur -- called an action scope, as it applies only on the action. The scope may define the starting boundary or the ending one (e.g., "after sailing termination", "before $<$B\_sig$>$ is True" in Fig.\ref{fig:reqEX}).

     Fig.\ref{fig:reqEX} presents the main variations for starting/ending a context (\textit{e.g.,} None, after operational constraint is true, until operational constraint becomes true or before operational constraint becomes true). Other alternatives can be expressed by the main variation. For example, "while R is true" can be expressed by after and until as "after R is true" and "until not R". It is worth noting that, "Before" and "Until" define the same end of the valid period which is "R is true". "Until" mandates the precondition(s)/action(s) to hold till "R is true", but "Before" does not care about their status. 
     \raisebox{-\totalheight}{\includegraphics[width=0.7\textwidth, height=3.3cm]{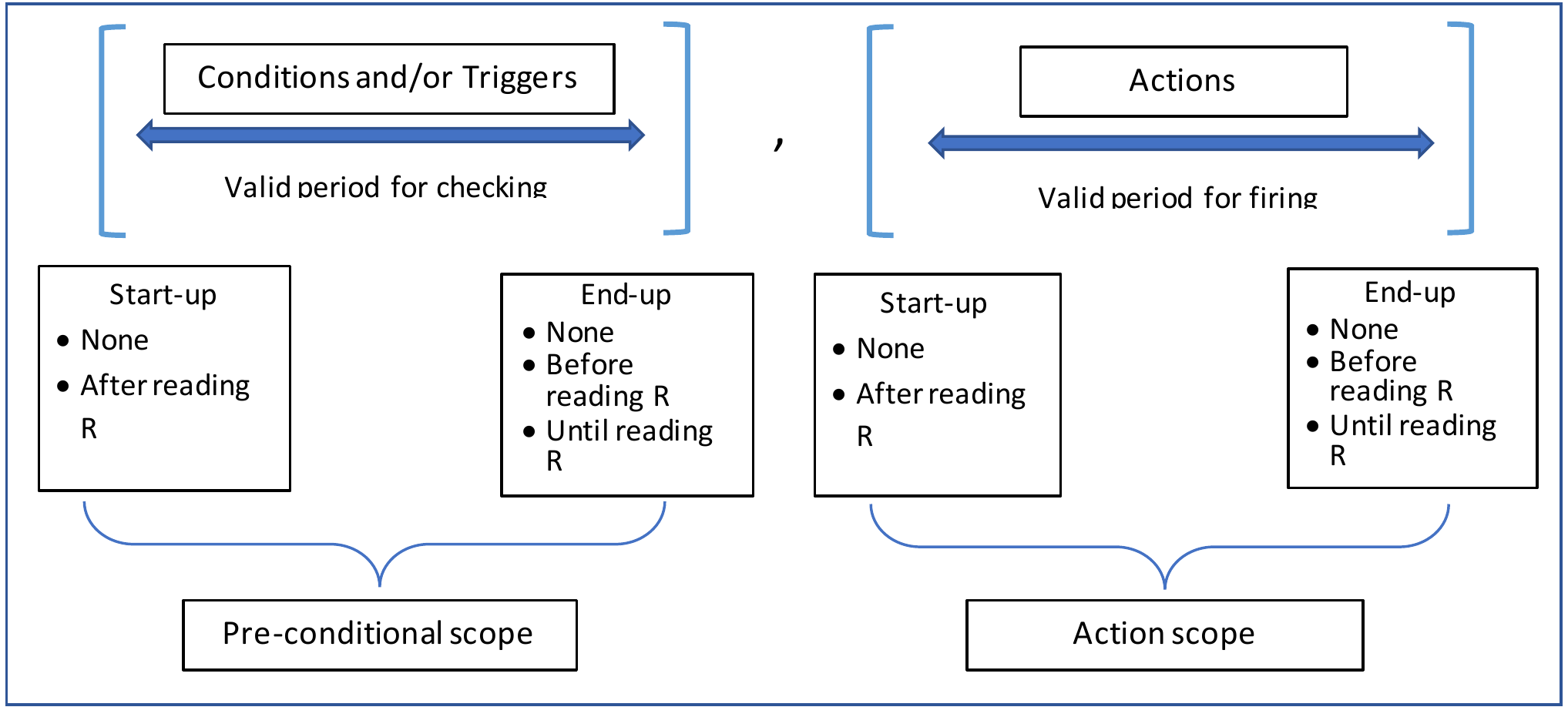} } \end{tabular}
     
    \\ \hline
    
    \multirow{4}{2cm}{}
   % &\textbf{Core-segment} & is mandatory to all components types. The core-segment express the main portion of the component including: the operands, the operator and negation flag/property (e.g., in "if X exceeds 1" the "X" and "1" are the operands and "exceeds" is the operator in the semi-formal semantic and "$>$" is the operator in the formal semantic). It could be mapped later to a proposition in temporal logic as indicated in Fig.\ref{fig:RCMUML}. \\ \cline{2-3}
    &  \textbf{Valid-time}& \begin{tabular}{p{14cm}} represent the valid time period of the given component (e.g., in "the vehicle warns the driver by acoustical signals $<E>$ for 1 second" the action is hold for 1 second length of time \cite{houdek2013system}). Valid-time can be a part of any component. \end{tabular}\\ \cline{2-3}
  \multirow{3}{2cm}{\rot{\textbf{Sub-Component}}}  &\textbf{Pre-elapsed-time} & \begin{tabular}{p{14cm}} is the consumed time length from an offset point --before an action to occur or a condition to be checked (e.g., "After less than 2 seconds" in Fig.\ref{fig:reqEX}). This type is only eligible to action and condition components. \end{tabular}\\ \cline{2-3}
    
    & \textbf{In-between-time}& \begin{tabular}{p{14cm}} express the length of time between two consecutive events to occur in the repetition case (e.g., "every 1 seconds" in Fig.\ref{fig:reqEX}). Such sub-component type is eligible to action and trigger components as indicated in Fig.\ref{fig:RCMUML}.\end{tabular} \\ \cline{2-3}
    
   & \textbf{Hidden constraint}& \begin{tabular}{p{14cm}} allows an explicit constraint to be defined on a specific operand within a component. For example, in "if the camera recognizes the lights of an advancing vehicle, \ul{the high beam headlight} \textbf{that is activated} is reduced to low beam headlight within 5 second"\cite{houdek2013system}. The \textbf{that is activated} is a constraint defined on the operand \textit{the high beam headlight}). \end{tabular}%To store this information without loss, RCM stores the hidden constraint inside the relevant operand object as indicated in Fig.\ref{fig:RCMUML}. This structure is intrinsic for allowing the nested hidden constraints. For example, "\textit{the entry of A1 \textbf{whose index is larger than \ul{the first value in A2}} \ul{that is larger than S1} shall be set to 0"}.\\\hline
   \\\hline

\end{tabular}
}
\label{tab:definitions}
\end{center}
\end{table*}
\section{\uppercase{Requirement Capturing Model}}\label{RCMOverview}

In this section, we present the details of the Requirements Capture Model (RCM). First, we explain the process we followed to develop the RCM. Then, we describe the RCM Model and the key components of the model. Third, we describe how to transform RCM to formal notations. Finally, we describe how to construct RCM from input textual requirements. It is important to note that this last step is out of the scope of this project, and thus we do not include all the details of the extraction process and we do not cover the evaluation of the extraction algorithm in this paper.

\subsection{RCM Development Process}
To identify the key requirement properties we needed to support in a generic reference model for safety-critical requirements, we reviewed a large number of natural language-based critical system requirements collected from many sources: \cite{jeannet2016debugging, thyssen2013behavioral, fifarek2017spear, lucio2017ears, dick2017requirements, bitsch2001safety, teige2016universal, lucio2017EARSAnalysis, mavin2009EARS, ACE, rolland1992natural, macias1995method} and 15 requirement representation approaches listed in Table.\ref{tab1:approaches}.

\begin{figure}[h]
	\centering
	\includegraphics[width=\columnwidth,height=.06\textheight]{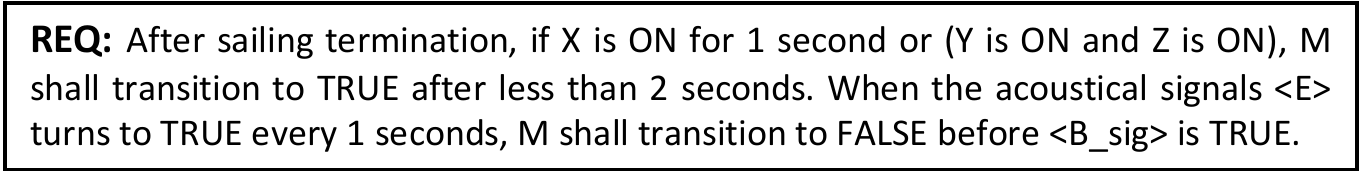}
	\caption{Crafted multi-sentence requirement "REQ"}
	\label{fig:EX}
\end{figure}

We identified 19 distinct properties that we grouped into 8 abstract properties (4 components and 4 sub-components). These are listed with their description in Table.\ref{tab:definitions}. Fig.\ref{fig:EX} shows a manually crafted example requirement that reflects most of these components and sub-components used through the properties description for a better understanding.

%provides the list and description of 11 key properties we identified categorised as 4 components and 7 sub-components.
%A: action, C: condition, and T: trigger(event). In real-time systems, we identified a suite of time-related properties including: SP: Scope of a pre-condition Startup-phase, (6) EP: Scope of a pre-cond Endup-Phase, (7) SA: Scope of an action  Startup-phase, (8)	EA: Scope of an action Endup-phase, (9) vt: valid-time attached to each component type (e.g., A-vt indicate the valid time of action components), (10) pt: pre-elapsed-time attached to each component types (e.g., A-pt indicates the pre-elapsed-time of the action components), and (11) rt: in-between-time attached to the eligible component types similarly to pre-elapsed-time. (4) hidden: Hidden-constraint,% \todo{ please fix this paragraph, also if A1, A2.. are not used anywhere else then delete them and just keep BTC, EARS, ECA.... we already have defined these in the related work section}

\begin{table*}[!t]
\caption{Exisiting approaches proposed properties and Supported formats}
\small Properties Codes $\rightarrow$ \textbf{A:}action  /	\textbf{C:}condtion	/ \textbf{T:}trigger /	\textbf{hidden:}Hidden-constraint /
\textbf{SP:}pre-cond Startup-phase /	\textbf{EP:}pre-cond Endup-Phase /
\textbf{SA:}action  Startup-phase /	\textbf{EA:}action Endup-phase /
\textbf{vt:}valid-time /	\textbf{pt:}pre-elapsed-time    /  \textbf{rt:}in-between-time

%\begin{center}
\scalebox{0.85}{

\begin{tabular}{|p{.5cm}|p{3.5cm}|p{.2cm}|p{.2cm}|p{.2cm}|p{.2cm}|p{.2cm}|p{.2cm}|p{.2cm}|p{.2cm} |p{.2cm}|p{.2cm}|p{.25cm}|p{.5cm}|p{.25cm} |p{.5cm}|p{.25cm}|p{.5cm}|p{.25cm}|p{.5cm}|p{.75cm}| }
\hline
\multicolumn{2}{|c|}{Approach} &\multicolumn{19}{|c|}{Requirement properties} \\\hline%\\cline{3-21}

\multicolumn{1}{|c}{}& \multicolumn{1}{|c|}{} &      \multicolumn{4}{|c|}{Action}	&\multicolumn{3}{|c|}{Condition}	&	\multicolumn{3}{|c|}{Trigger}	&	\multicolumn{8}{|c|}{Req-Scope} & \\\cline{3-20}	

%\multicolumn{1}{|c}{}&\multicolumn{1}{c|}{}&\multicolumn{1}{c}{}&\multicolumn{1}{c}{}&\multicolumn{1}{c}{}&\multicolumn{1}{c|}{}&\multicolumn{1}{c}{}&\multicolumn{1}{c}{}&\multicolumn{1}{c|}{}&\multicolumn{1}{c}{}&\multicolumn{1}{c}{}&\multicolumn{1}{c}{}&\multicolumn{2}{|c|}{SP}&\multicolumn{2}{|c|}{EP}&\multicolumn{2}{|c|}{SA}&\multicolumn{2}{|c|}{EA}& \\\cline{1-20}
%Code & Source & core-seg &	Vt &	Rt &	Pt &	core-seg &	Vt & Pt	 &core-seg & Vt &	Rt &  core-seg &	Vt & core-seg &	vt &  core-seg &	Vt & core-seg &	Vt &	Hidden \\\hline

Code & Source & A &	A-vt &	A-rt &	A-pt &	C &	C-vt & C-pt	 & T & T-vt &	T-rt &  SP &	SP-vt & EP & EP-vt &  SA &	SA-vt & EA & EA-vt &	Hidden \\\hline

%A1 & BTC &	1&	1&	&1	&1	&1	&	&	&	&	&1	&	&	&	&1 & 1	&1	&1 &\\\hline
A1	&\multirow{2}{3.5cm}{BTC \cite{BTCTh, teige2016universal}}&1	&1	&	&1	&1	&1	&	&	&	&	&1	&	&	&	&1	&1	&1	&1	& \\\cline{3-21}
	&	    &1	&1	&	&1	&	&	&	&1	&1	&	&1	&	&	&	&1	&1	&1	&1	& \\\hline
A2	&\multirow{3}{3cm}{EARS \cite{mavin2009EARS}}	&1	&	&	&	&1	&	&	&1	&	&	&	&	&	&	&	&	&	&	& \\\cline{3-21}
	&   	&1	&	&	&	&1	&	&	&	&	&	&1	&	&	&	&	&	&	&	& \\\cline{3-21}
	&   	&1	&	&	&	&	&	&	&1	&	&	&1	&	&	&	&	&	&	&	& \\\hline
A3	& EARS-CTRL	\cite{lucio2017EARSAnalysis}&1	&	&	&	&	&	&	&1	&	&	&1	&	&	&	&	&	&1	&	& \\\hline
A4	&ECA \cite{van2010atomate}	&1	&	&	&1	&1	&1	&1	&1	&1	&1	&1	&	&	&	&	&	&	&	& \\\hline
A5	&\multirow{2}{3.5cm}{boilerplates \cite{rupp2009rupboilerplates}}	&1	&	&	&	&1	&	&	&	&	&	&	&	&	&	&	&	&	&	& \\\cline{3-21}
	&	    &1	&	&	&	&	&	&	&1	&	&	&	&	&	&	&	&	&	&	& \\\hline
A6	&\multirow{3}{3cm}{Safety templates \cite{fu2017generic}}	&1	&	&1	&1	&	&	&	&	&	&	&	&	&	&	&1	&	&1	&	& \\\cline{3-21}
	&   	&1	&	&1	&1	&1	&	&	&	&	&	&	&	&	&	&1	&	&1	&	& \\\cline{3-21}
	&	    &1	&	&	&	&	&	&	&1	&	&1	&1	&	&1	&	&	&	&	&	& \\\hline
A7	&Req Lang	\cite{marko2015combining}&1	&1	&	&	&1	&	&	&1	&	&	&1	&	&1	&	&	&	&	&	& \\\hline
A8	&\multirow{4}{3cm}{CFG \cite{scott2004context,CFGTh}}	&1	&	&	&	&1	&	&	&	&	&	&	&	&	&	&	&	&	&	&1 \\\cline{3-21}
	&   	&1	&	&	&	&	&	&	&1	&	&	&	&	&	&	&	&	&	&	&1 \\\cline{3-21}
	&   	&1	&	&	&	&	&	&	&	&	&	&	&	&	&	&1	&	&	&	&1 \\\cline{3-21}
	&   	&1	&	&	&	&	&	&	&	&	&	&	&	&	&	&	&	&1	&	&1 \\\hline
A9	&ACE \cite{ACE}	&1	&	&	&	&1	&	&	&	&	&	&	&	&	&	&	&	&	&	&1 \\\hline
A10	&PENG \cite{schwitter2002english}	&1	&	&	&	&1	&	&	&1	&	&	&1	&	&1	&	&1	&	&1	&	&1 \\\hline
A11	&Structured English \cite{yan2015formal}	&1	&	&	&1	&1	&	&	&1	&	&	&1	&	&1	&	&1	&	&1	&	&  \\\hline
A12	&TBNLS \cite{esser2007obtaining}	&1	&1	&	&	&1	&1	&	&	&	&	&1	&	&1	&	&	&	&1	&	&  \\\hline
A13	&\multirow{6}{3cm}{Real-time \cite{konrad2005realTime}}	&1	&	&	&1	&	&	&	&	&	&	&	&	&	&	&1	&	&1	&	& \\\cline{3-21}
	&   	&1	&	&1	&	&	&	&	&	&	&	&	&	&	&	&1	&	&1	&	& \\\cline{3-21}
	&   	&1	&1	&	&	&	&	&	&	&	&	&	&	&	&	&1	&	&1	&	& \\\cline{3-21}
	&   	&1	&	&1	&	&	&1	&	&	&	&	&1	&	&1	&	&	&	&	&	& \\\cline{3-21}
	&   	&1	&1	&	&	&	&1	&	&	&	&	&1	&	&1	&	&	&	&	&	& \\\cline{3-21}
	&   	&1	&	&	&	&1	&	&	&	&	&	&1	&	&1	&	&	&	&1	&	& \\\hline
A14	&Dawyer \cite{dwyer1999patterns}	&1	&	&	&	&1	&	&	&	&	&	&1	&	&1	&	&	&	&1	&	& \\\hline
A15	&Pattern\_based Req	 \cite{berger2019multiple} &1	&1	&	&1	&1	&1	&	&	&	&	&	&	&	&	&	&	&	&	& \\\hline
%A16	&RCM &1	&1	&1	&1	&1	&1	&1	&1	&1	&1	&1	&1	&1	&1	&1	&1	&1	&1	&1 \\\hline
\end{tabular}
\label{tab1:approaches} }
%\end{center}
\end{table*}

We then analysed 15 of the existing approaches (outlined in the related work section) against these 19 requirement properties as presented in Table.\ref{tab1:approaches}. The approaches (rows) are encoded A1 to A15, and requirement properties are encoded as columns. An approach can be represented in more than one row. This reflects that some approaches might support multiple properties, but these properties cannot be used in the same requirement -- the template or pattern does not support having certain properties in one requirement. The cell value equals "1" if the property is supported in this template. 

This table does not reflect the limitations/restrictions that these approaches apply on a given property formatting or order - i.e. condition must come before action, or scope comes before condition. Our analysis of this table illustrates that: (1) no approach covers all requirement properties possibly because this would make it too complex to use; (2) almost all approaches support action components as a core element; (3) approaches A1: Btc\cite{BTCTh} and A11: Structured-English\cite{konrad2005realTime} are the most expressive approaches as they cover majority of the properties; and (4) the valid-time property for the StartUp and the EndUP phases of the pre-conditional scope is not supported by any of these approaches although its appearance in the analysed requirements. 

\begin{figure*}[!t]
	\centering
	\includegraphics[width=\textwidth, height=.35\textheight]{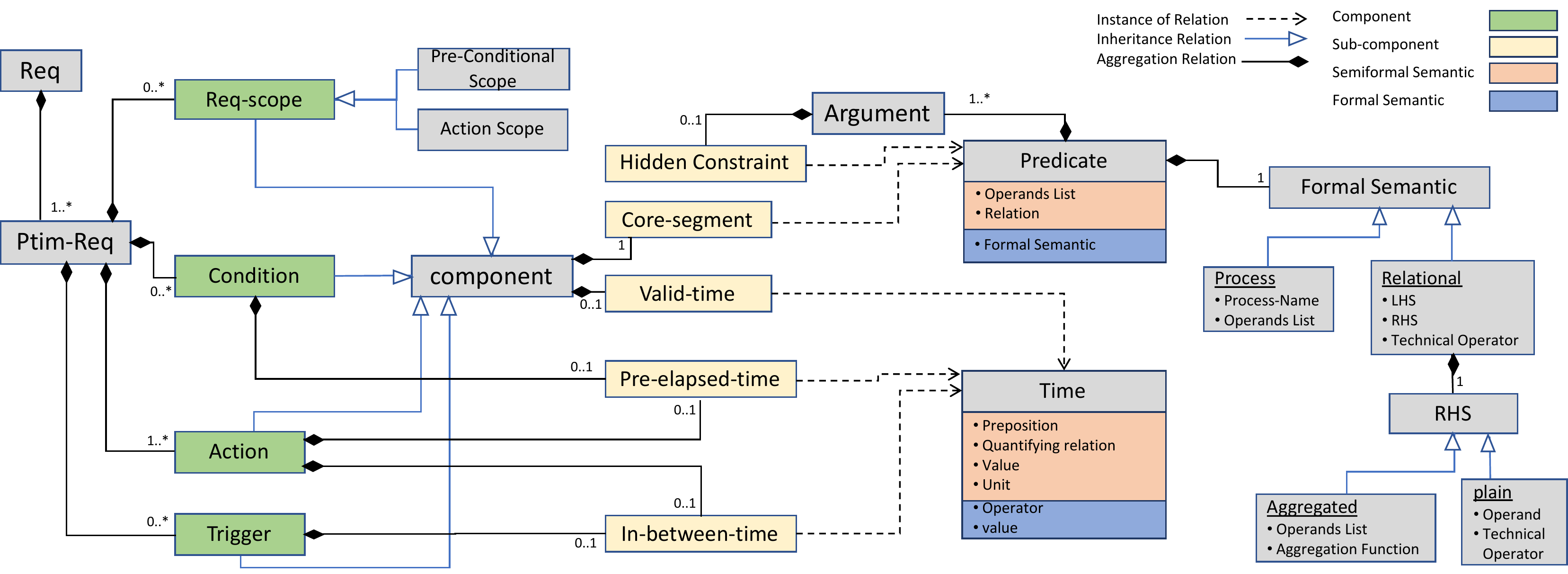}
	\caption{RCM meta-model (simplified)}
	\label{fig:RCMUML}
\end{figure*}

%As shown in Fig.\ref{fig:reqEX}, REQ has two primitive requirements PR[1] and PR[2]. PR[1] = "After sailing termination, if X exceeds 1 for 1 second or (Y is ON and Z is ON), M shall transition to TRUE after less than 2 seconds." and PR[2] = "When the acoustical signals <E> turns to TRUE every 1 seconds, M shall transition to FALSE before <B\_sig> is TRUE.". Components of each primitive requirement are presented in separate blocks. In each block, sub-components are separated by horizontal line. Fig.\ref{fig:reqEX} presents the RCM representation of the REQ example highlighting the encapsulation of semi-formal semantic (in black) and formal semantic (in red). It also provides the MTL representation of each  primitive requirement, see subSection.\ref{RCMFormalization}.2.

\subsection{RCM Domain Model} \label{RCMModel}
%Fig.\ref{fig:RCMUML} presents a detailed meta-model of the RCM. A system is represented as a set of requirements R. Each requirement $R_{i}$, may have one or more primitive requirements PR where \{$R_{i}$ = $<$ $PR_{n}$ $>$ and n$>$0\}. Each $PR_{j}$ represents only one requirement sentence. In the REQ example, we have two primitive requirements PR[1] and PR[2] as it contains two sentences as indicated in Fig.\ref{fig:reqEX}. 

The RCM is designed to capture the requirements properties listed above while relaxing the ordering and formatting restrictions presented by the existing techniques. In RCM, a system is represented as a set of requirements R. Each requirement $R_{i}$ represented by one RCM and may have one or more primitive requirements PR where \{$R_{i}$ = $<$ $PR_{n}$ $>$ and n$>$0\}. Each $PR_{j}$ represents only one requirement sentence, and may include condition(s), trigger(s), action(s) and requirement scope(s). The detailed meta-model structure of the RCM to one requirement $R_{i}$ is presented in Fig.\ref{fig:RCMUML}.

%\todo{we need to talk about some adjustment to this model asap - can we remove valid-time from component and treat it as the other types of time? also why condition does not have inbetweentime?  precond and action scope are realisation of the Req-scope? how would this work if i have action scope condition action trigger? what does grey means in the figure? why does predict compose argument which compose hidden constraint?}

The figure shows that a primitive \emph{requirement} is composed of four requirement \emph{component} types: \emph{condition}, \emph{trigger}, \emph{action} and \emph{requirement scope}. Except for action(s), the existence of each of these components is optional in a primitive requirement. A requirement component has a \emph{component core-segment} that expresses the main portion of the component, and optionally could also have a \emph{valid-time}: the component's valid time-length. The \emph{pre-elapsed-time} sub-component can only appear with a condition or action component. An \emph{in-between-time} sub-component can only appear with Trigger or Action components according to the reviewed scenarios (e.g., requirements and representation formats). A \emph{hidden-constraint} is an optional sub-component to an operand. To store this information without loss, RCM stores the hidden constraint inside the relevant operand object as indicated in Fig.\ref{fig:RCMUML}. This structure is intrinsic to allow the nested hidden constraints. For example, "\textit{the entry of A1 \textbf{whose index is larger than \ul{the first value in A2}} \ul{that is larger than S1} shall be set to 0"}. 

All five sub-components are instances of either \emph{Predicate} or \emph{Time} structure. The \emph{Predicate} structure consists of the operands, the operator and negation flag/property (e.g., in "if X exceeds 1" the "X" and "1" are the operands and "exceeds" is the operator in the semi-formal semantic and "$>$" is the operator in the formal semantic). The \emph{Time} structure stores the unit, value and quantifying relation (e.g., "for less than 2 seconds", "2" and "seconds" are the unit and value respectively, "less than" is the semi-formal quantifying relation whose formal semantic is "$<$"). Since the \emph{Predicate} and \emph{Time} structures are the infrastructure of the entire properties, they are designed to encapsulate the semi-formal and formal semantic allowing mappability to multiple TL. The details of formal semantic are described in section.\ref{FSSection}.

\begin{figure*}[!t]
	\centering
	\includegraphics[width=\textwidth,height=.57\textheight]{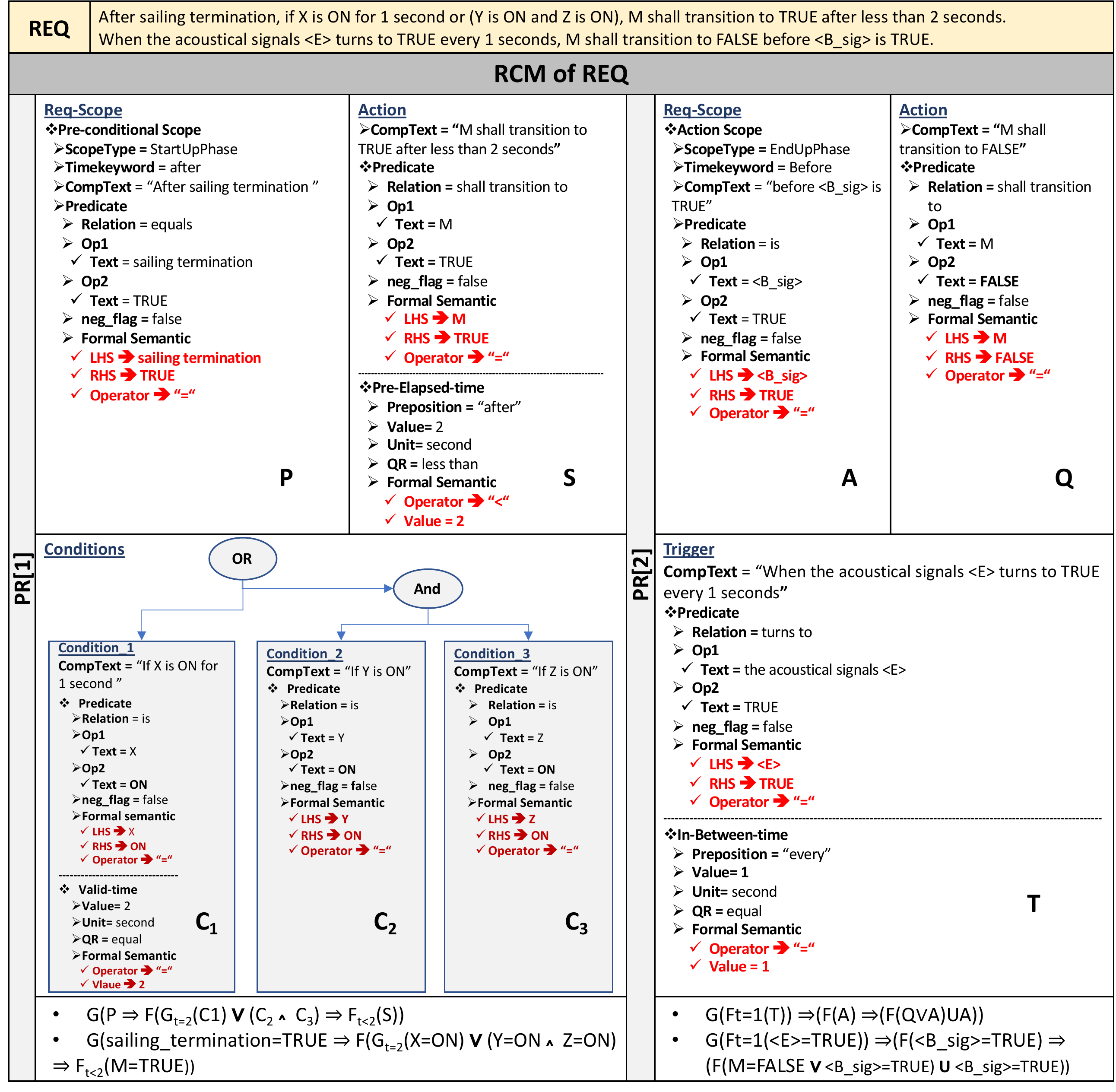}
	\caption{An example presents multi-sentence requirement "REQ" and the corresponding RCM representation}
	\label{fig:reqEX}
\end{figure*}

Components with the same type can be stored as a tree --the most suitable to keep nested relation appropriately, where leafs are the components, and inner nodes are coordinating relationships (e.g., check the conditions components of PR[1] in Fig.\ref{fig:reqEX}). 

%\todo{@Aya, the formal semantic part is not described here. If you plan to describe it later, please metion here that the remaining formal semantics components are described in section ABC}

%Table.\ref{tab:definitions} provides the detailed description of the identified properties (components and sub-components). For a better understanding, we define a crafted requirement - REQ example in \ref{fig:reqEX}  combining variations of common critical system requirements properties - and  use  it  throughout the explanation in Table.\ref{tab:definitions}. As shown in Fig.\ref{fig:reqEX}, REQ has two primitive requirements PR[1] and PR[2]. PR[1] = "After sailing termination, if X exceeds 1 for 1 second or (Y is ON and Z is ON), M shall transition to TRUE after less than 2 seconds." and PR[2] = "When the acoustical signals <E> turns to TRUE every 1 seconds, M shall transition to FALSE before <B\_sig> is TRUE.". Components of each primitive requirement are presented in separate blocks. In each block, sub-components are separated by horizontal line. Fig.\ref{fig:reqEX} presents the RCM representation of the REQ example highlighting the encapsulation of semi-formal semantic (in black) and formal semantic (in red). It also provides the MTL representation of each  primitive requirement, see subSection.\ref{RCMFormalization}.2.

%Fig.\ref{fig:reqEX}, shows the RCM representation of the REQ example, section.\ref{construction} illustrate the construction step by step. The figure also provides the MTL representation of each primitive requirement, see subSection.\ref{RCMFormalization}.%\todo{@Aya, please add descripotion of each part in the figure here}

\textbf{Example RCM:}
Fig.\ref{fig:reqEX} shows the RCM representation of the REQ example. REQ has two primitive requirements. PR[1] constituting of five components \{"After sailing termination", "if X is ON for 1 second", "Y is ON", "Z is ON", and "M shall transition to TRUE after less than 2 seconds"\}. PR[2] has three components \{"When the acoustical signals $<$E$>$ turns to TRUE every 1 seconds", "M shall transition to FALSE", and "before $<$B\_sig$>$ is TRUE"\}. Components of each primitive requirement are presented in separate blocks in the figure. In each block, sub-components (predicates and time structures) are separated by horizontal line. The figure also highlights the encapsulation of semi-formal semantic (in black) and formal semantic (in red). Components with the same type (e.g., conditions in PR[1]) are represented by tree structure. In addition, the MTL representation of each primitive requirement is provided, see subSection.\ref{RCMFormalization}

\subsection{RCM Transformation}
%To illustrate how the RCM is transformed into 
In this section, we illustrate transformation into temporal logic (TL)- as an example of formal notations. We first illustrate: (1) the mapping between the RCM to TL, and (2) the formalization of the RCM infrastructure (i.e., Predicate and Time structures). Then, we provide the transformation process.%\todo{@Aya, we need to explain the mapping table, what does each row and col mean and how it works ; JG - agreed}

\begin{table*}[!b]
	\caption{RCM mapping to MTL \& CTL}
	\begin{center}
		\scalebox{0.875}{
	
		%\begin{tabular}{|l|ll|l|m{3cm}|m{3.5cm}|m{1cm}|}
		\begin{tabular}{|lp{1cm}|l|m{3cm}|m{1.8cm}|m{2.5cm}|m{2.7cm}|m{.55cm}|}
			
			\hline
			\multicolumn{5}{|c|}{\textbf{RCM}} & \multicolumn{3}{c|}{\multirow{1}{2cm}{\textbf{TL Mapping}}} \\ \cline{1-8}
			\multicolumn{2}{|p{3.5cm}|}{\textbf{Properties (component/ subcomponents)}} & \multicolumn{2}{c|}{\textbf{Versions}} & \textbf{Applicable on}&  \multicolumn{1}{c|}{MTL}& \multicolumn{1}{c|}{CTL}& \\
			\hline
			
			\multicolumn{2}{|l|}{Action} & 1 & A: do something &&  A& A & \\\cline{1-7}
			
			\multicolumn{1}{|l}{\multirow{3}{1.5cm}{Pre-condition}} 
			&\multicolumn{1}{|p{2cm}|}{Condition} &2&If S & \multirow{3}{2cm}{Action \textit{(P in mapping)}} & {$G(S$ $\implies$ $P)$} &{$AG(S$ $\implies$ $P)$}&  \multirow{9}{.5cm}{\rot{Temporal Modality}}\\ \cline{2-4}\cline{6-7}
		    &\multicolumn{1}{|p{2cm}|}{Trigger}&3&When S & & {$G(S$ $\implies$ $P)$} & {$AG(S$ $\implies$ $P)$}&\\ \cline{2-4}\cline{6-7}
		    &\multicolumn{1}{|p{2cm}|}{Conditions and triggers}&4&When S, IF Q & & {$G((S$ $\wedge$ Q) $\implies$ $P)$} &{$AG((S$ $\wedge$ Q) $\implies$ $P)$}&\\\cline{1-7}

			\multirow{5}{2cm}{Req-Scope: (Preconditional-Scope / Action-Scope)} &  \multicolumn{1}{|p{1.5cm}|}{\multirow{1}{*}{StartUP}} & 5& After S & \multirow{5}{2.2cm}{ Pre\-condition/ action \textit{(P in mapping)}} & {$G(S$ $\implies$ $F(P))$} &    {$AG(S$ $\implies$ $AG(AF(P)))$} &  \\\cline{2-4} \cline{6-7}
			& \multicolumn{1}{|p{1.5cm}|}{\multirow{2}{*}{EndUP}} &6&Before S & & {$F(S)$ $\implies$ $(F(P \lor S) \textbf{U} S)$} & $A[((AF(P \lor S)) \lor AG(\neg S)) \textbf{\emph{W}} S]$ &\\ \cline{3-4}\cline{6-7}
			&                        \multicolumn{1}{|c|}{}    & 7&Until S &&  {$F(P) U S$} & {$AF(P) U S$}&\\ \cline{2-4}\cline{6-7}
			& \multicolumn{1}{|p{2cm}|}{\multirow{2}{2cm}{StartUP and EndUp}} &8&After Q \& Before S Between Q and S && {$G((Q\wedge\neg S \wedge F(S))$ $\implies$ $F(P \lor S) \textbf{U} S))$} &  $AG((Q \wedge \neg S) \implies A[((AF(P \lor S) \lor AG(\neg S)) \textbf{W} S])$&\\ \cline{3-4}\cline{6-7}
			&   \multicolumn{1}{|c|}{}&9& After Q Until S \& While Z \{Q=Z\&S=$\neg$ Z\} && {$G((Q\wedge\neg S)$ $\implies$ $F(P \textbf{U} S))$} & $AG((Q \wedge \neg S) \implies A[(AF(P \lor S) \textbf{W} S])$&\\ \cline{1-4}\cline{5-8}

			 \multicolumn{2}{|c|}{\multirow{3}{*}{Pre-elapsed-time}} &10&After c time & \multirow{3}{2cm}{Condition/ Action \textit{(P in mapping)}}& {$F_{t=c}(P)$} && \multirow{9}{.5cm}{\rot{Time notation}}\\ \cline{3-4} \cline{6-6}
			& &11&after at-most c time && {$F_{t\leq c}(P)$} &&\\ \cline{3-4} \cline{6-6}
			& &12&after at-least c time && {$F_{t\geq c}(P)$} &&\\ \cline{3-4} \cline{6-6}
			& &13&after less-than c time && {$F_{t< c}(P)$} &&\\ \cline{3-4} \cline{6-6}
			& &14&after greater-than c  && {$F_{t> c}(P)$} &&\\ \cline{1-6}
			
			\multicolumn{2}{|c|}{\multirow{2}{*}{Validation-time}} & 15&for c time & \multirow{3}{2cm}{Condition/ Trigger/ Action \textit{(P in mapping)}}& {$G_{t=c}(P)$} &&\\ \cline{3-4} \cline{6-6}
			& &16&for at-most c time && {$G_{t\leq c}(P)$} &&\\ \cline{3-4} \cline{6-6}
			& &17&for at-least c time && {$G_{t\geq c}(P)$} &&\\ \cline{3-4} \cline{6-6}
			& &18&for less-than c time && {$G_{t< c}(P)$} &&\\ \cline{3-4} \cline{6-6}
			& &19&for greater-than c  && {$G_{t> c}(P)$} &&\\ \cline{1-6}
			
			\multicolumn{2}{|c|}{\multirow{3}{*}{In-between-time}} & 20&every c time & \multirow{3}{2cm}{Action/ Trigger \textit{(P in mapping)}} & {$G(F_{t=c}(P))$} &&\\ \cline{3-4} \cline{6-6}
			& & 21&every at-most c time  && {$G(F_{t\leq c}(P))$} &&\\ \cline{3-4} \cline{6-6}
			& & 22&every at-least c time  && {$G(F_{t\geq c}(P))$} &&\\ \cline{3-4} \cline{6-6}
			& & 23&every less-than c time && {$G(F_{t< c}(P))$} &&\\ \cline{3-4} \cline{6-6}
			& & 24&every greater-than c  && {$G(F_{t> c}(P))$} &&\\ \hline
		    \multicolumn{2}{|p{3cm}|}{Hidden-Constraint} &25& Whose S &  \multirow{2}{2cm}{P is Any component} &&{$AG(\exists S$ $\implies$ $P)$}&bran\\ 
			 & &&&  &&&ching\\ \hline

		\end{tabular}}
		\label{tab1:RCMtoMTL}
	\end{center}
\end{table*}

\subsubsection{RCM and Temporal Logic} \label{mappingSec}
In order to formally model a given requirement represented by RCM in temporal logic (TL), we have to define a set of transformation rules. A TL formula $F_{i}$ is built from a finite set of proposition variables AP by making use of boolean connectives (e.g., "AND", "OR") and the temporal modalities (e.g., U (until)) \cite{LTLCTL2015, brunello2019synthesis}. Within such formula, each proposition letter is expressed by a true/false statement and may be attached with time notation in some versions of temporal logic (e.g., MTL). Consider the following sentence:"After the button is pressed, the light will turn red until the elevator arrives at the floor and the doors open\cite{brunello2019synthesis}". Such sentence can be captured by the following TL formula: 

$p \implies (q U(s\wedge v))$ 

where p, q, s, and v are proposition variables corresponding to "the button being pressed", "the light turning red", "the elevator arriving", and "the doors opening", respectively. \\

We use the following to build the mapping between RCM and TL:
\begin{enumerate}
    \item \textbf{Propositions and time notations}: Given that, RCM components and sub-components are expressed as predicates or time structures as indicated in Fig.\ref{fig:RCMUML}. These structures are eventually mapped to proposition and time notations in the corresponding temporal logic formula (e.g., the action component "\textbf{M shall transition to TRUE} \ul{after less than 2 seconds}" mapped to "$F_{t<2}$(S)", where S and "t$<$2" represent the predicate in bold and time phrase underlined).
    \item \textbf{Coordinating relations}: The booleans connecting propositions can be obtained from coordinating relations connecting multiple components with the same types. Such relations are represented by tree for each component type as discussed before (e.g., the condition components "X is ON for 1 second or (Y is ON and Z is ON)" mapped to "$(G_{t=2}(C1)\lor(C2\wedge C3))$". 
    \item \textbf{Temporal modality}: The temporal modalities can be identified based on the component type (e.g., the type of the component "After sailing termination" is "pre-conditional-scope startup-phase" mapped to "$\Longrightarrow$" 
\end{enumerate}

%\todo{rcm to mtl and ctl table is not described as how we developed what it means how it works etc.
%after we do so we need to then refer to the example RCM and explain how using these mapping and primitive reqs 1 and 2 we managed to generate the notations}

To demonstrate the robustness of the RCM and capability to transform to different formal notations, we provide here a mapping into two examples of temporal logic, Metric Temporal Logic (MTL) \cite{alur1993real, koymans1990specifying} and CTL \cite{clarke2008design}, as shown in Table \ref{tab1:RCMtoMTL} as a proof of multiple map-ability. We chose these notations as they are widely used in model checking as indicated in \cite{konur2013survey} and \cite{frappier2010comparison, LTLCTL2015} respectively. We base our temporal-modality and time-notation mapping on the mapping done in \cite{konrad2005realTime}. 

The first column in the Table.\ref{tab1:RCMtoMTL} shows the RCM properties (components and sub-components) employed in formal roles, each attached with alternatives if any (e.g., The pre-conditions may be conditions, triggers, or both of them based on the given requirement). Possible structures corresponding to each property version are listed in the third column (i.e., the used keywords (e.g., when) are just examples, any replaceable keyword could be used). The fourth column indicates which components can be linked to each property type. The MTL and CTL representations of each property are presented in the fifth and sixth columns respectively, where these notations are grouped based on their formal types in the last column.

MTL is a real-time extension of linear-time temporal logic (LTL) \cite{szalas1995temporal}. It has time notations and temporal operator. MTL consists of propositional variables, logical operator (\textit{e.g.,} $\neg$, $\lor$, $\wedge$ and $\implies$), temporal operators (\textit{e.g.,} until $U_{I}$) where \textit{I} is an interval of time. In addition, MTL has a timed-version of always, eventually operators. However, it doesn't handle point intervals (\textit{e.g.,} [$t_{1}$, $t_{1}$] as indicated in \cite{konur2013survey}). MTL assumes the existence of external and discrete clock updates constantly (fictitious-clock model as discussed in \cite{alur1991techniques}). 

Complementary to MTL that models linear systems, CTL model the system in a tree-like structure (i.e, there are multiple paths, any one of them might be an actual path that is realized) but it does not support time notations. In CTL, If $f_{1}$ and $f_{2}$ are formulas, then so are $\neg f_{1}$, $f_{1}$$\wedge$ $f_{2}$, $f_{1}$$\lor$ $f_{2}$, AG$f_{1}$, EG$f_{1}$, A[ $f_{1}$ U  $f_{2}$], and E[$f_{1}$ U  $f_{2}$], were, A and E means "along All paths (Inevitably)" and "there-exist one path (possibly)" respectively. CTL could determine if a given artifact possesses safety properties (e.g., "all possible executions of a program avoid undesirable condition"). 

%\todo{@Aya, please include summary for CTL too here}
%\todo{@Aya, what does this paper say that you based your work on?}

%RCM can be also mapped into more abstract temporal logic as (\textit{e.g.,} LTL \cite{szalas1995temporal}).

\subsubsection{RCM and Formal Semantics}\label{FSSection}
%\textcolor{red}{any better title} //updated
%The basic units of the RCM are predicate and time structures that are mapable to temporal logic. Temporal logic has various versions each with small differences. In order to support transformation into multiple versions of temporal logic with minimal adjustment in the parsing technique, RCM encapsulates formal semantics with semiformal semantics within predicate and time structures, as indicated in Fig.\ref{fig:RCMUML}. The formal semantics of a predicate covers three formats:
Temporal logic has multiple versions exhibiting slight differences. In order to support the transformation to multiple versions with minimal adjustment in the transformation technique, RCM encapsulates formal semantics with semi-formal semantics. Design-wise, RCM augments the formal semantic in the basic units, predicate and time structures in Fig.\ref{fig:RCMUML}, that are mapable to temporal logic, as indicated in the previous subsubsection. The formal semantic of a predicate covers three formats:

\begin{itemize}
    \item \textbf{Process format}: is suitable to predicates express functions or process (e.g., "the monitor sends a request $REQ\_Sig$ to the station" $\longrightarrow $ "send(the$\_$monitor, the\_station,$REQ\_Sig$)"). 
    
    \item \textbf{Relational format with plain RHS}: the type is suitable for assignment predicates (e.g., "set X to True" $\longrightarrow $"X = True"), comparison predicates (e.g., "If X exceeds Y" $\longrightarrow $"X $>$ Y") and changing state predicates (e.g., "the window shall be moving up" $\longrightarrow $"the\_window = moving-UP").
    
    \item \textbf{Relational format with aggregated RHS}: this format is similar to the previous one but the RHS is expressed with aggregating function (e.g., "If the fuel level is less than the min value of Thr1 and Thr2" $\longrightarrow $"the\_fuel\_level $<$ "min(Thr1, Thr2)").
\end{itemize}

Similarly, the formal semantic is added to time structure in which the technical time operator (e.g., \{$>,<,=,\leqslant,\geqslant$\}) is identified (e.g., "for at least 2 seconds" $\longrightarrow $ "t $\geqslant$ 2").

\begin{figure*}[!t]
%.7
\scalebox{.94}{
%.46
\begin{minipage}{.65\textwidth}
    %\begin{algorithm}
    \captionof{algorithm}{{RCM-to-MTL Transformation}}\label{transformation}
    %\captionof{algorithmic}{\textbf{Alg1: RCM-to-MTL Transformation}}\label{transformation}
\begin{algorithmic}[1]
\small
    \State \textbf{Input: } 
                   \Statex[1] R: RCM-to-MTL indexed Mapping Rules 
                   \Statex[1] PrimReq: primitive requirement of interest
    \State \textbf{Output: }
                   \Statex[1] mTLFormula: generated Formula
    \Procedure{}{}
    \State \textbf{Step 1:} Prepare each component
            \ForAll{comp $\in$ PrimReq}
                \Statex[2] Comp.Formal $\gets$ Comp.CoreSegment.getFormalSemantic()
               %\Statex[3].getFormalSemantic()
                \ForAll{timeInfo $\in$ comp}
                \Statex[3]  Comp.Formal $\gets$ comp.AttachTimeSemantic(timeInfo, R\{10:24\})
                %\Statex[3] AttachTimeSemantic(timeInfo, R\{9:25\})
                %\Statex[3](timeInfo, R\{9:25\})
                \EndFor
            \EndFor    

    \State \textbf{Step 2:} Aggregate components of the same type
            
            \ForAll{compTree $\in$ CompTypeTree}
                 \Statex[2] aggVal $\gets$ aggRel(compTree)
              
                 \Procedure{aggRel}{Tree compTree}
                    \If{compTree is leaf}
                         \Statex[4] return compTree.data.Formal;
                    \Else 
                    
                       \Statex[4]  return  "(" +  aggRel(compTree.Left) 
                          \Statex[4]       + compTree.data.LR() +
                       \Statex[4]         aggRel(CompTree.Right) ")”
                    \EndIf
                \EndProcedure

                \Statex[2] map.put(compTree.Type, aggVal)
            \EndFor
     
    \State \textbf{Step 3: }Prepare Preconditions
            \Statex[1] preConds $\gets$ preparePrecond(map[Triggers], map[Conditions], R\{2:4\})
            %\Statex[1] preparePrecond(map[Triggers], map[Conditions], R{6:8})

    \State \textbf{Step 4: } Prepare LHS and RHS
            \Statex[1] lHS $\gets$ prepareSide(preConds, R\{5:9\})
            \Statex[1] rHS $\gets$ prepareSide(Actions, R\{5:9\})

    \State \textbf{Step 5: }Generate Formulat
            \If{lHS $\neq$ $\phi$}
               \Statex[2] mTLFormula $\gets$ lHS + "$\longrightarrow$" + rHS
            \Else
               \Statex[2] mTLFormula $\gets$ rHS
            \EndIf

           \Statex[1] return mTLFormula
    \EndProcedure
    \end{algorithmic}
    
    %\end{algorithm}
\end{minipage}%
%.25
\begin{minipage}[h]{.3\textwidth}
    \centering
    \includegraphics[height=.63\textheight]
    {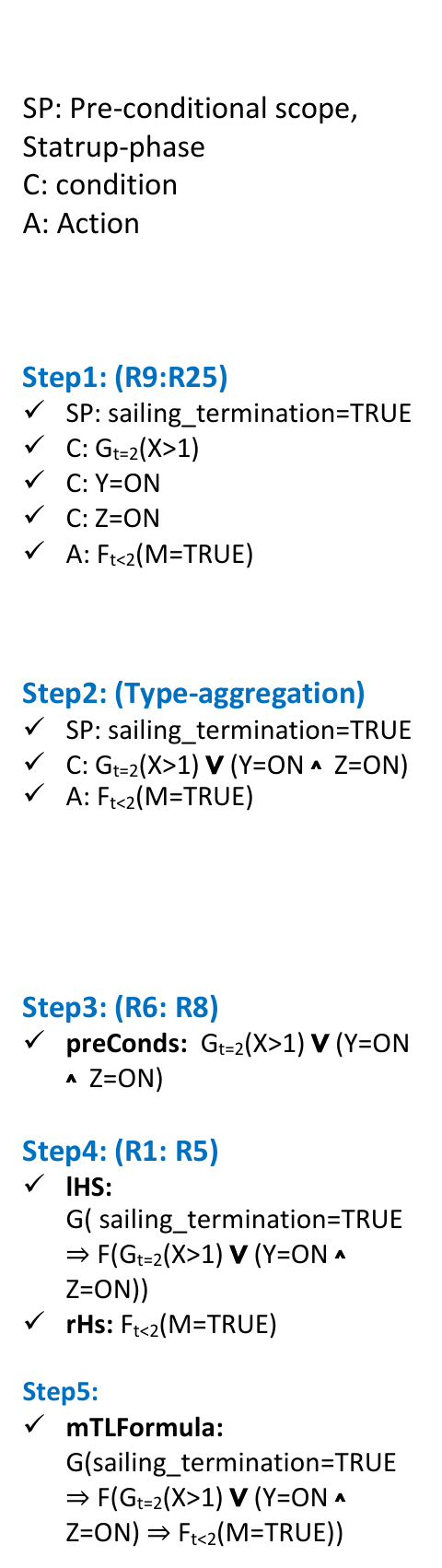}
    \captionof{figure}{Step by Step generation of PR[1] from Fig.\ref{fig:reqEX}}  \label{formal}
  
\end{minipage}
}
\end{figure*}

%\todo{should we move section 5.2 here and merge these together?}

\subsubsection{RCM Transformation Algorithm}\label{RCMFormalization}
%In Table \ref{tab1:RCMtoMTL}, we summarise the meta-models mapping between RCM and MTL.
To accomplish the automatic transformation from RCM-to-MTL, we use the mapping rules provided in Table \ref{tab1:RCMtoMTL} on the obtained formal semantics of the given primitive requirements. Algorithm \ref{transformation} shows the automatic transformation pseudo-code annotated in Fig.\ref{formal} with each step output for PR[1] in the REQ Fig.\ref{fig:reqEX}. 

First, we get the formal semantics of each component according to Subsection.\ref{FSSection}. Then, we compute the formal semantics of the entire tree (i.e., leaf nodes represent components and inner nodes represent logical relations as discussed before) of each component type through the recursive function aggRel. After that, we construct the main parts of the formula (i.e., preCondtions, LHS and RHS) in Step3 and 4 with the help of RCM-to-MTL mapping rules in Table \ref{tab1:RCMtoMTL}. Finally, we generate the entire formula based on the bound sides either "LHS $\longrightarrow$ RHS" or "RHS" as in Step5.

%============================================================================================================================

\begin{comment}

We have developed a generic \emph{Requirement Capturing Model (RCM)}. RCM is a more comprehensive approach to representing NL-based requirement for real-time critical systems. RCM supports translating these common representations into a variety of formal requirement modelling notations for specialised automated requirements verification.

Within the frameworks of existing solutions a format is defined based on the structure of the entire requirement sentence. Thus, these approaches are sensitive to restructure-based paraphrasing. Conversely, our RCM overcomes this constraint by holding the exiting key elements of the requirement independent on their order and occurrence within the source natural language requirement sentence. Another advantage is that RCM supports storing a detailed level of information (at a formal level) enabling a direct transformation into different formal notations (e.g. temporal logic versions in our case) for further formal verification. RCM can also be used to provide a refined and unified version of the requirements in the form of constrained natural language CNL (textual writing is the most popular form of requirements representation among stakeholders \cite{kocerka2018analysing}), for validation purposes. %The refined requirements ease the validation process by users as this process can be only done manually.

\end{comment}

%\import{./}{Formalisation}
\section{\uppercase{Evaluation}} \label{evaluation}
\subsection{Benchmark Datasets}

\begin{comment}
We evaluate the coverage of our proposed RCM on 5 curated data-sets. These datasets were extracted from existing case studies in the literature \todo{ @Aya, please rewrite these sentences could not understand including (1) expressing proposed templates/defined formats on representing requirements in different domains considering different writing styles (BTC-DS and Aggregated-DS), (2) evaluating formalization approaches (Arsenal-DS and CARA-DS) and (3) extracted from an online available system requirements that are not tweaked for any special use (CruiseControl-DS). }
The characteristics of the included requirements from these data-sets are summarised in Table\ref{tab1:DataSets}. 

\begin{table}[htbp]
\caption{Data-Sets Description}
\begin{center}
\begin{tabular}{|m{2.5cm}|m{2cm}|m{1cm}|m{1.3cm}|}
\hline
   \textbf{Dataset} & \textbf{Sources} & \textbf{$\#$Req} & \textbf{$\#$Sentences}\\
\hline
Arsenal-DS  & \cite{ghosh2016arsenal} & 20 & 20 \\
\hline
CARA-DS  & \cite{yan2015formal} & 8 & 8 \\
\hline
%BTC-DS  & \cite{BTCTh} & 30 & 30 \\
BTC-DS & \cite{BTCTh} & 33 & 33 \\ %excess words
\hline
CruiseControl-DS  &  \cite{houdek2013system} & 35 & 43 \\
\hline
Aggregated-DS  &  \cite{jeannet2016debugging}\cite{ thyssen2013behavioral}, \cite{fifarek2017spear}, \cite{lucio2017ears}, \cite{dick2017requirements}, \cite{bitsch2001safety}, \cite{teige2016universal}, \cite{lucio2017EARSAnalysis}, \cite{mavin2009EARS}, \cite{ACE} & 57 & 58 \\
\hline

\textbf{Total} &   &  153 & 162 \\
\hline
\end{tabular}
\label{tab1:DataSets}
\end{center}
\end{table}
\end{comment}

We evaluate the coverage of our proposed RCM on 162 requirement sentences. These requirements were extracted from existing case studies in the literature and grouped into three sub datasets as follows: (1) expressiveness dataset (81 requirements): these are requirements collected from papers that introduced different requirement templates and formats in different domains and considering different writing styles in \cite{BTCTh} \cite{jeannet2016debugging}\cite{ thyssen2013behavioral}, \cite{fifarek2017spear}, \cite{lucio2017ears}, \cite{dick2017requirements}, \cite{bitsch2001safety}, \cite{teige2016universal}, \cite{lucio2017EARSAnalysis}, \cite{mavin2009EARS}, \cite{ACE}, (2) formalisation dataset (28 requirements): these are requirements extracted from papers that introduced requirement formalisation techniques including \cite{ghosh2016arsenal, yan2015formal} with total of 28 requirements and (3) online sources (43 requirements): these are requirements extracted from an online available critical-system requirements including \cite{houdek2013system}. These requirements are available from \footnote{\label{Ev} Dataset:\url{https://github.com/ABC-7/RCM-Model/tree/master/dataSet}}.

\begin{figure}[htbp]
\includegraphics[width=\columnwidth, height = .256\textheight]{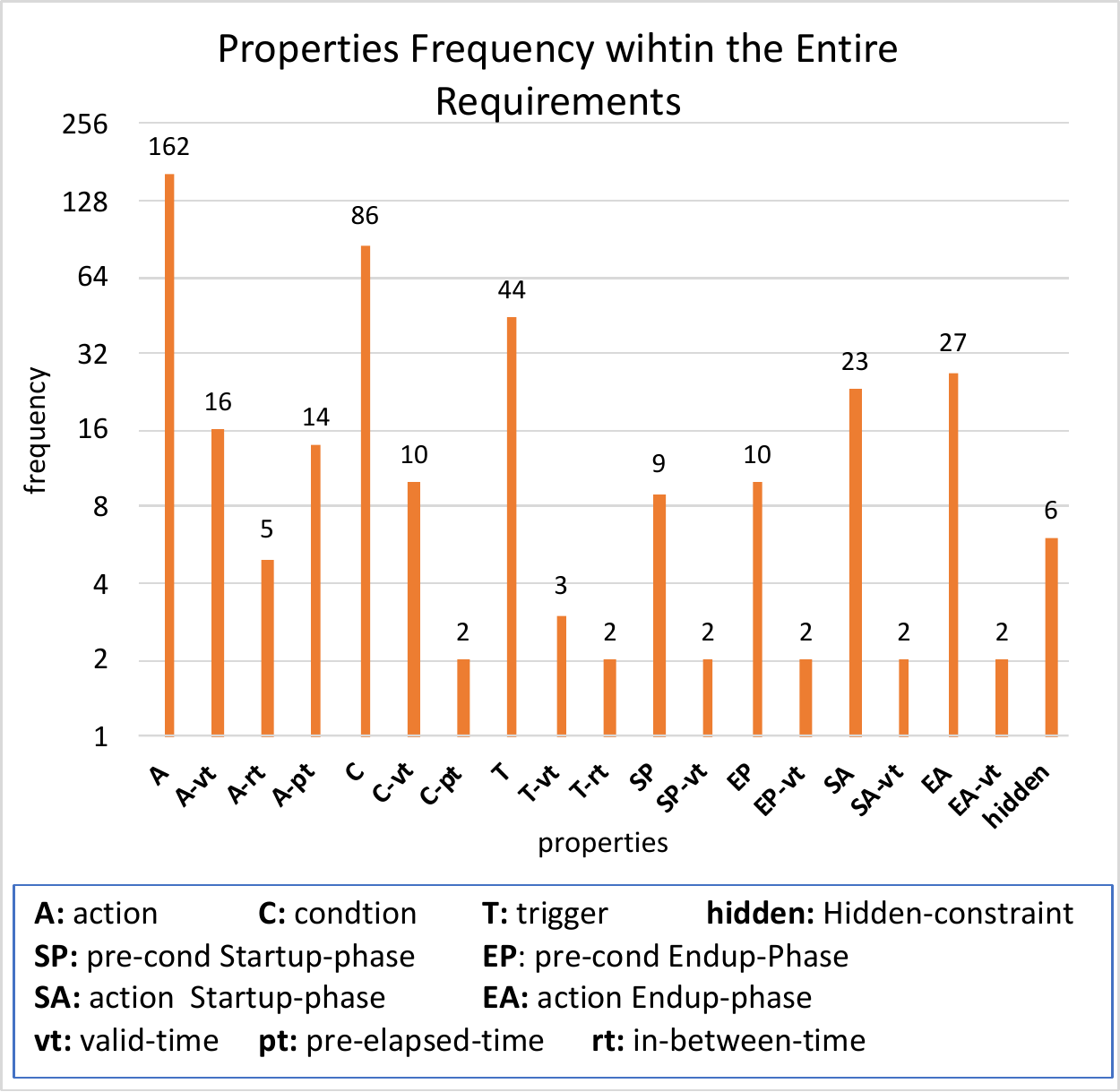}
\caption{Properties' Frequency wihtin the Entire Requirements}

\label{fig:dis}
\end{figure}

%of the 19 properties presented in Table.\ref{tab1:approaches}. 
Fig.\ref{fig:dis} presents the percentages of each of the 19 requirement properties (components/sub-components) within the entire dataset. The figure shows that time-based and hidden constraints existed in a few requirements compared to the key requirement components such as action, trigger, and condition. Overall, the distribution of the properties is biased towards the popular properties that exist in most approaches.% in Table.\ref{tab1:PropToApp}.

%The 11 possible properties are: (1) A: action (core-segment), (2) C: condition (core-segment), (3) T: trigger (core-segment), (4) hidden: Hidden-constraint, (5) SP: Scope of a pre-condition Startup-phase, (6) EP: Scope of a pre-cond Endup-Phase, (7) SA: Scope of an action  Startup-phase, (8)	EA: Scope of an action Endup-phase, (9) vt: valid-time attached to each component type (e.g., A-vt indicate the valid time of action components), (10) pt: pre-elapsed-time attached to each component types (e.g., A-pt indicates the pre-elapsed-time of the action components), and (11) rt: in-between-time attached to the eligible component types similarly to pre-elapsed-time. 

\begin{figure}[htbp]
\includegraphics[width=\columnwidth, height= .17\textheight]{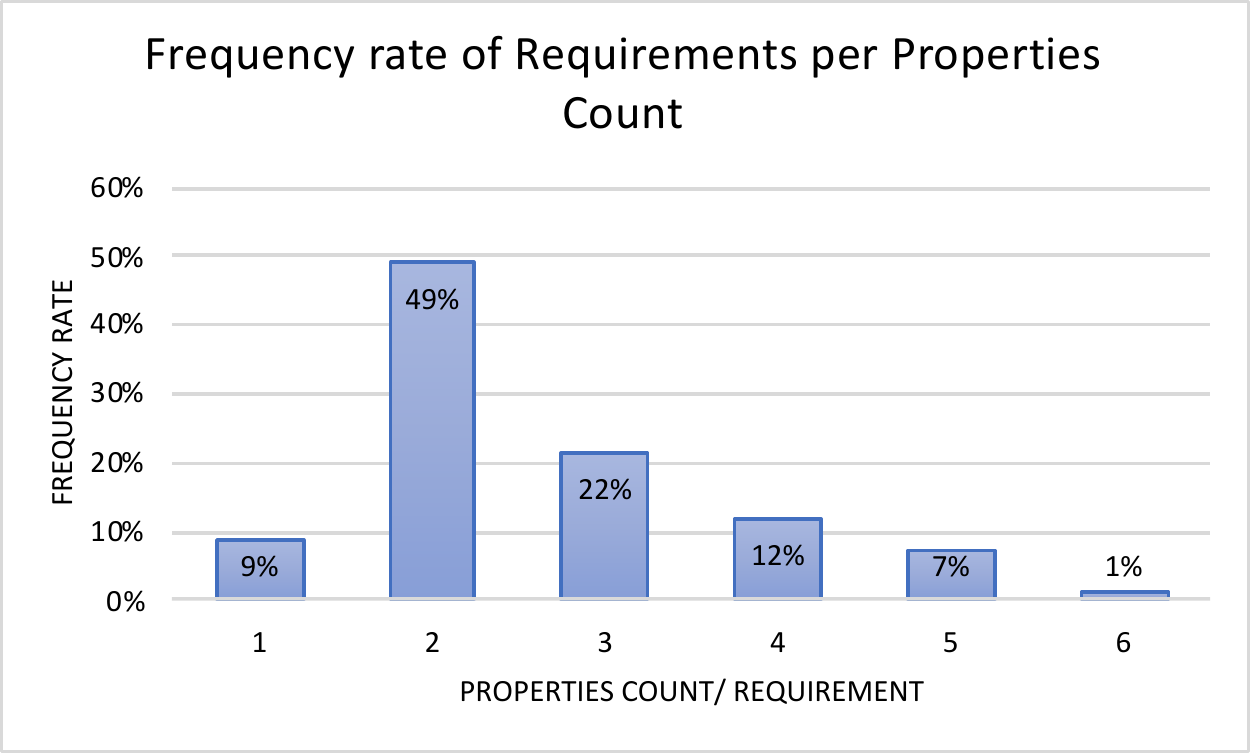}
\caption{Frequency rate of Requirements per Properties Count  }
\label{fig:freq}
\end{figure}

Fig.\ref{fig:freq} shows the relative complexity of the 162 requirements. We grouped the requirements based on the count of their existing properties (i.e., number of properties per requirement increases $\uparrow$, its complexity increases $\uparrow$). The following examples show two requirements with one and six properties respectively, where each property is separately underlined: (1) "\ul{the monitor mode shall be initialized to INIT}", and  (2) "\ul{after X becomes TRUE} \ul{for 2 seconds}, \ul{when Z turns to 1} \ul{for 1 second}, \ul{Y shall be set to TRUE} \ul{every 2 seconds}". In Fig.\ref{fig:freq}, each group represents the count of properties regardless of the type of the property - i.e., R1: requirement with condition and action, and R2: requirement with trigger and action, both have 2 properties). For each group, we calculated the percentage of requirements. Fig.\ref{fig:freq} presents the properties count used for each requirements group on the x-axis and the corresponding requirements percentage on the y-axis. This shows that a large portion of the entire requirements sentences 9\%, 49\% and 22\%, only consist of one, two and three properties respectively. On the other hand, 20\% of the requirements sentences consist of more than three properties. %\todo{@Aya how come A only appeared 128 times in figure5??? it should be in all requirements? also do we have the combinations? for example action only, action and condition, action and trigger,...}
%\textcolor{red}{ give example of what 1 property would be, two properties could be, etc. Also does this mean you evaluated on simple requirementss only? how does this map to the coverage evaluation research question}

\subsection{Evaluation Experiments}

\begin{figure}[htbp]
\includegraphics[width=\columnwidth, height =.23\textheight]{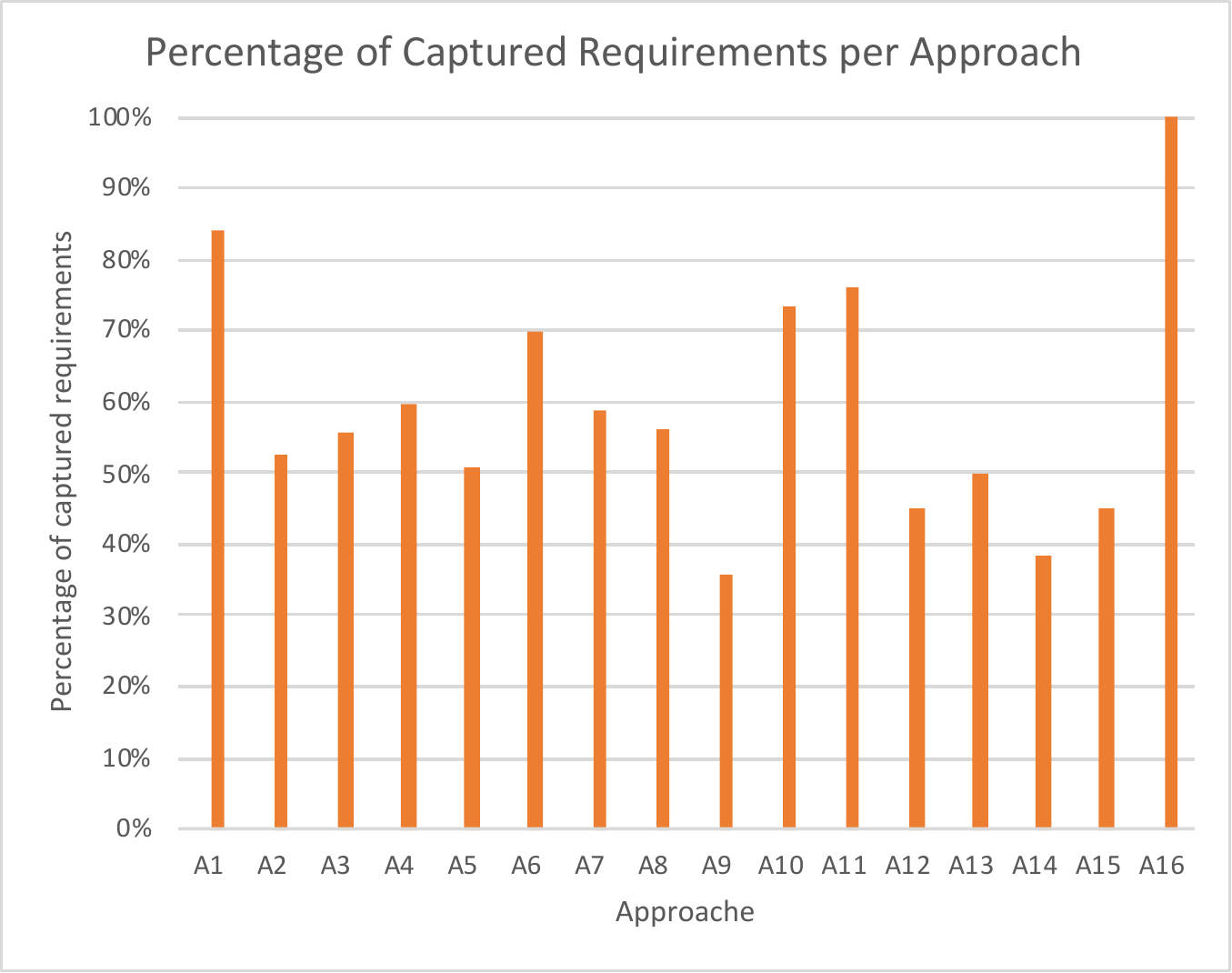}
\caption{Percentage of Captured Requirements per Approach {\small(RCM represented by A16 and the other represent by codes proposed in subsection.\ref{RCMModel})} }

\label{fig:app}
\end{figure}

\textbf{Experiment1. RCM expressiveness.} We evaluated our proposed RCM reference model's ability to capture and represent the requirements in our test dataset compared to 15 exiting approaches in Table.\ref{tab1:approaches}. To do this, we manually labelled all the requirements in the dataset against the 19 requirement properties we identified in section 4. After that, we wrote a script to check each requirement (identified properties) against all existing approaches to assess if the approach provides a boilerplate or a template that supports representing the requirement or not. The results are available online \footnote{Approaches representations, and evaluation: \url{https://github.com/ABC-7/RCM-Model/blob/master/Approaches-Evaluation.xlsx}}. Fig.\ref{fig:app} summarises the results of our analysis as percentage of the test requirements that each approach supports. 

This shows that none of the existing 15 approaches is able to represent the entire dataset of requirements. This is mainly for two reasons: (1) missing properties in the used templates e.g., A1 does not support StartUp-phase Pre-conditional scope (SP), or (2) restrictions on the included properties in a requirement format e.g., A2:EARS does not support the existence of the trigger (core-segment) and a ReqScope (core-segments) using the same format. In addition, $\approx$4\% of the test requirements were not covered by any of these approaches combined. An example is "if the maximum deceleration is [insufficient] before a collision with the vehicle ahead, the vehicle warns the driver by acoustical signals for 1 seconds every 2 seconds", where the existing properties are: condition (core-segment), StatrtUp-phase Pre-conditional scope (SP core-segment), action (core-segment), action valid-time (Vt) , and action in-between-time (Rt). These properties do not exist together in the same representation of any of the 15 approaches, see Table \ref{tab1:approaches}. 

In contrast, our proposed RCM requirements model can represent all of the 162 requirements sentences. This is because it covers all properties that exist in the other approaches and puts no restriction on the included properties in one requirement (i.e., any property could exist in the requirement format).

Existing approaches require extension in two cases: (1) considering new requirement properties, and (2) considering new formats i.e, defining a set of properties that can exist together in one format regulated by customized grammatical rules. In contrast, since RCM covers all properties of the other approaches and more and puts no constraints on properties used in requirement, it is powerful enough to represent all requirements that can be represented by all the other approaches. It can also be used in other scenarios not currently supported by any of the 15 approaches, due to the fact that it does not enforce any restriction on the input requirement formats. 

RCM does however have two main limitations: (1) it is designed for behavioral requirements of critical systems --based on the known templates,CNLs and formats used for formalization as illustrated in subsection.\ref{RCMModel}, and (2) it requires NL-extraction techniques i.e., the current NL-extraction processes primitive requirements expressed in one sentence. 
%A second example is "if the camera recognizes the lights of an advancing vehicle, the high beam headlight that is activated is reduced to low beam headlight within 5 seconds" including properties:  condition (core-segment), action (core-segment), hidden constraint, and action pre-elapsed-time (Pt). Similarly, these properties together only exist in the RCM row in table \ref{tab1:approaches}. 

\textbf{Experiment2: RCM to formal notations}
We applied our RCM-to-MTL and RCM-to-CTL transformation rules to the dataset of the 162 requirements. In this experiment, we used our NLP tool to extract RCM from the textual representation of the 162 requirements (out of scope of this paper). We then manually reviewed all the extracted RCM models, fixed all the broken RCM extractions manually. Once we had the full list of 162 RCM models, we applied the RCM-to-Formal transformation ruless as outlined in Section 4. We then manually reviewed all the generated formal notations. The full list of RCMs representation and the corresponding automatically generated MTL and CTL formulas is available online \footnote{RCM-Representation and formal-notation: \url{https://github.com/ABC-7/RCM-Model/tree/master/RCM-Auto-Transformation}}. 

We successfully transformed 156 out of the 162 requirement RCM models into MTL notations. The other 6 requirements were partially correct. These 6 requirements turned out to involve hidden constraints expressed with $\exists$ and $\forall$ properties with a branching structure that is not supported by MTL, since it is linear. For example, the requirement "the cognitive threshold of a human observer shall be set to a deviation that is less than 5. \cite{houdek2013system}" was correctly represented in RCM, but the generated MTL is partially correct "G(the cognitive threshold of a human observer = the deviation)". The correct generation should be "AG(($\exists$ deviation$<$5) $\implies$ (the cognitive threshold of a human observer = deviation))" provided by CTL. 
 
Similarly, CTL could represent requirements with hidden constraints correctly, but it provides partial solutions for requirements with time notation e.g., validation-time, pre-elapsed-time and in-between-time. In total, it is capable of representing 120 requirements correctly and provides partial solutions 42 ones due the inclusion of time notation (e.g., the requirement "if air\_ok signal is low, auto control mode is terminated within 3 sec" has a partially correct generated CTL formal "AG([air\_ok signal = low] $\implies$ [auto control mode.crrStatus = terminated])", but the correct formula should be G([air\_ok signal = low] $\implies$ [F{t=3}(auto control mode.crrStatus = terminated)]) provided by MTL.

%\textcolor{red}{Table V, I could not undersstand the multiline per group of tool(s)}

%By aggregating the properties distribution and the complexity analysis we can conclude that most of the requirements sentences could be captured by the existing approaches. 

%Fig. shows the percentage of requirements sentences captured by each approach of the 15 approach.

%Since the RCM is designed to represent functional/behavioral requirements, it may not fit for the other types of requirements. In addition, complex requirements may require breakdowns at first. For-example, the complex requirement "the traffic lights of the main road and of the side road must not display a green signal at the same time" could not be directly represented by the RCM. At first, it should be broken down into: (1) "the main road shall display a green signal unless the side road displays a green signal" and (2) "the side road shall display a green signal unless the main road displays a green signal". After that, the RCM could represent each as a primitive requirement consisting of action and condition components. 

\section{\uppercase{Summary}} \label{conclusion}
We introduced a new requirements capturing model - RCM - that represents an abstract and intermediate representation of safety-critical system requirements. RCM defines a wide range of key requirement elements and attributes that may exist in an input requirement. The model allows for standardising the textual requirements extraction process, simplifies the transformation rules to convert requirements to formal notations, and more importantly avoids re-writing existing requirements which is a complex and error prone task. We compared the coverage of our RCM model to 15 existing requirements modelling approaches using 162 diverse requirements curated from the literature. Our results show that RCM can capture a wider range of requirements compared to others due to its separation of concerns of source natural language requirements input format from reference model representation. In addition, we provided a suite of RCM-to-MTL transformation rules and presented the corresponding automatically generated MTL representation of the evaluation dataset. For our future work, we are developing an automated requirements extraction technique to populate RCM from textual requirements in addition to requirements quality checking and visualising tool of the system RCM models. 

%We are also working on a suite of transformation rules that can be used to convert RCM-based requirements into different formal notation. 

%\textcolor{red}{please include summary of the evaluation and findings...}
%RCM copes with the nature of natural language requirements differentiating between dependencies among multiple requirements sentences and dependencies among multiple requirements.
%\textcolor{red}{details of the evaluation}

\bibliographystyle{apalike}
{\small
\bibliography{ref}}

\end{document}